\documentclass[a4paper,fleqn,english]{article} 
\usepackage{latexsym}\usepackage{graphicx}\usepackage{babel}
\usepackage{epsfig}
\begin{document}
\catcode`\@=11
\@addtoreset{equation}{section}
\def\theequation{\arabic{section}.\arabic{equation}}
\def\appendix{\renewcommand{\thesection}{\Alph{section}}\setcounter{section}{0}
 \renewcommand{\theequation}
 {\mbox{\Alph{section}.\arabic{equation}}}\setcounter{equation}{0}}
\def\schw{Schwarzschild }
\def\maketitle{\thispagestyle{empty}\setcounter{page}0\newpage
 \renewcommand{\thefootnote}{\arabic{footnote}}
 \setcounter{footnote}0}
\renewcommand{\thanks}[1]{\renewcommand{\thefootnote}{\fnsymbol{footnote}}
 \footnote{#1}\renewcommand{\thefootnote}{\arabic{footnote}}}
\newcommand{\preprint}[1]{\hfill{\sl preprint - #1}\par\bigskip\par\rm}
\renewcommand{\title}[1]{\begin{center}\Large\bf #1\end{center}\rm\par\bigskip}
\renewcommand{\author}[1]{\begin{center}\Large #1\end{center}}
\newcommand{\address}[1]{\begin{center}\large #1\end{center}}
\newcommand{\pacs}[1]{\smallskip\noindent{\sl PACS numbers:
 \hspace{0.3cm}#1}\par\bigskip\rm}
\def\babs{\hrule\par\begin{description}\item{Abstract: }\it} 
\def\eabs{\par\end{description}\hrule\par\medskip\rm}
\renewcommand{\date}[1]{\par\bigskip\par\sl\hfill #1\par\medskip\par\rm}
\newcommand{\ack}[1]{\par\section*{Acknowledgments} #1} 

\def\dinfn{Dipartimento di Fisica, Universit\`a di Trento\\ 
 and Istituto Nazionale di Fisica Nucleare,\\
 Gruppo Collegato di Trento, Italia \medskip}

\def\csic{Consejo Superior de Investigaciones Cient\'{\i}ficas \\
Instituto de Ciencias del Espacio (ICE/CSIC)\\
 Campus UAB, Facultat Ciencies, Torre C5-Parell-2a planta, \\
08193 Bellaterra (Barcelona), Spain \medskip}

\def\ieec{Institut d'Estudis Espacials de Catalunya (IEEC), \\
Edifici Nexus, Gran Capit\`a 2-4, 08034 Barcelona, Spain \medskip}

\def\icrea{$^d$ Instituci\`o Catalana de Recerca i Estudis Avan\c{c}ats 
(ICREA), Barcelona, Spain\medskip}

\def\guido{Guido Cognola\thanks{e-mail: \sl cognola@science.unitn.it\rm}}
\def\sergio{Sergio Zerbini\thanks{e-mail: \sl zerbini@science.unitn.it\rm}}
\def\luciano{Luciano Vanzo\thanks{e-mail: \sl vanzo@science.unitn.it\rm}}
\def\emilio{Emilio Elizalde\thanks{e-mail: \sl elizalde@ieec.uab.es\rm}}
\def\sergei{Sergei D.~Odintsov\thanks{e-mail: \sl odintsov@ieec.fcr.es\rm}}


\newcommand{\s}[1]{\section{#1}}
\renewcommand{\ss}[1]{\subsection{#1}}
\newcommand{\sss}[1]{\subsubsection{#1}}
\def\M{{\cal M}} 
\newcommand{\ca}[1]{{\cal #1}} 
\def\segue{\qquad\Longrightarrow\qquad} 
\def\prece{\qquad\Longleftarrow\qquad} 
\def\hs{\qquad} 
\def\nn{\nonumber} 
\def\beq{\begin{eqnarray}} 
\def\eeq{\end{eqnarray}} 
\def\ap{\left.} 
\def\at{\left(} 
\def\aq{\left[} 
\def\ag{\left\{} 
\def\cp{\right.} 
\def\ct{\right)} 
\def\cq{\right]} 
\def\cg{\right\}} 
\newtheorem{theorem}{Theorem} 
\newtheorem{lemma}{Lemma} 
\newtheorem{proposition}{Proposition} 
\def\R{{\hbox{{\rm I}\kern-.2em\hbox{\rm R}}}} 
\def\H{{\hbox{{\rm I}\kern-.2em\hbox{\rm H}}}} 
\def\N{{\hbox{{\rm I}\kern-.2em\hbox{\rm N}}}} 
\def\C{{\ \hbox{{\rm I}\kern-.6em\hbox{\bf C}}}} 
\def\Z{{\hbox{{\rm Z}\kern-.4em\hbox{\rm Z}}}} 
\def\ii{\infty} 
\def\X{\times\,} 
\newcommand{\bin}[2]{\left(\begin{matrix}{#1\cr #2\cr}
 \end{matrix}\right)} 
\newcommand{\fr}[2]{\mbox{$\frac{#1}{#2}$}} 
\def\Det{\mathop{\rm Det}\nolimits} 
\def\tr{\mathop{\rm tr}\nolimits} 
\def\Tr{\mathop{\rm Tr}\nolimits} 
\def\rot{\mathop{\rm rot}\nolimits} 
\def\PP{\mathop{\rm PP}\nolimits} 
\def\Res{\mathop{\rm Res}\nolimits} 
\def\res{\mathop{\rm res}\nolimits} 
\renewcommand{\Re}{\mathop{\rm Re}\nolimits} 
\renewcommand{\Im}{\mathop{\rm Im}\nolimits} 
\def\dir{/\kern-.7em D\,} 
\def\lap{\Delta\,} 
\def\arccosh{\mbox{arccosh}\:} 
\def\arcsinh{\mbox{arcsinh}\:} 
\def\arctanh{\mboz{arctanh}\:} 
\def\arccoth{\mbox{arccoth}\:} 

\def\d{${}^-\kern-.9em\partial\,$} 
\def\al{\alpha}
\def\be{\beta}
\def\ga{\gamma}
\def\de{\delta}
\def\ep{\varepsilon}
\def\ze{\zeta}
\def\io{\iota}
\def\ka{\kappa}
\def\la{\lambda}
\def\ro{\varrho}
\def\si{\sigma}
\def\om{\omega}
\def\ph{\varphi}
\def\th{\theta}
\def\te{\vartheta}
\def\up{\upsilon}
\def\Ga{\Gamma}
\def\De{\Delta}
\def\La{\Lambda}
\def\Si{\Sigma}
\def\Om{\Omega}
\def\Te{\Theta}
\def\Th{\Theta}
\def\Up{\Upsilon}

\title{Homogeneous cosmologies in generalized modified gravity}

\author{Guido Cognola\thanks{cognola@science.unitn.it}, 
Sergio Zerbini\thanks{zerbini@science.unitn.it}}
\address{Dipartimento di Fisica, Universit\`a di Trento \\ 
and Istituto Nazionale di Fisica Nucleare \\ 
Gruppo Collegato di Trento, Italia}

\begin{abstract}
Dynamical system methods are used in the study of the 
stability of spatially flat homogeneous
cosmologies within a large class of generalized modified gravity models 
in the presence of a relativistic matter-radiation fluid. 
The present approach can be considered as the generalization 
of previous works in which only $F(R)$ cases were considered. 
Models described by an arbitrary function of all possible 
geometric invariants are investigated and general equations 
giving all critical points are derived. 

\end{abstract}


\section{Introduction}

Recent cosmological data support the fact that there 
is a strong evidence for a late accelerated expansion of the observable 
universe, apparently due to the presence of an effective positive and small 
cosmological constant of unknown origin. This is known as dark energy 
issue (see for example \cite{padmanabhan}). 
 
Modified gravity models are possible realizations 
of dark energy (for a recent review and alternative approaches 
see \cite{CST,rev3}), which may offer a quite natural 
geometrical approach again in the spirit of the original Einstein theory of
gravitation.
In fact, the main idea underlying these approaches to dark
energy puzzle is quite simple and consists in
adding to the gravitational Einstein-Hilbert action other gravitational terms 
which may dominate the cosmological evolution
during the very early or the very late universe epochs, 
but in such a way that General Relativity remains valid at intermediate epochs
and also at non cosmological scales. They are generalization of the 
$\Lambda$CDM model, namely Einstein gravity plus a 
positive cosmological constant, the simplest model for dark energy, which
however has to be confronted with well known difficulties, 
among them, the cosmological constant problem, an unsolved issue so far. 

In the present paper, first of all we shall generalise 
the analysis presented in 
\cite{monica}, where the stability of the de Sitter solution
(the vacuum invariant submanifold) has been investigated in a 
class of modified gravitational cosmological models 
defined in a Friedmann-Robertson-Walker (FRW) spatially flat space-time. The 
method is well known and consists in rewriting the equations of motion 
as a system of first-order
autonomous differential equations and makes use of the theory of dynamical
systems (see \cite{ellis,amendola,amendola1,dunsby,fay} and 
references therein). 
We remind that the stability or instability issue is really 
relevant in cosmology. 
For example, in the $\Lambda$CDM model it ensures that no future 
singularities will be present in the solution.
Within cosmological models, the stability or instability around a 
solution is of interest at early and also at late times. 

Then we will introduce also matter/radiation in the model and we 
shall generalise the method given in Ref.~\cite{amendola}, which permits
to determine all critical points of a $F(R)$ model.
Ordinary matter is important in reconstructing the expansion 
history of the Universe and probing the phenomenological relevance of the 
models (see for example the recent papers 
\cite{amendola,amendola1,dunsby,fay}, where the $F(R)$ case has been 
discussed in detail). 
Our generalisation consists in the extension of that method 
in order to include all 
possible geometrical invariants. This means that $F$ 
could be a generic scalar function of curvature, Ricci and 
Riemann tensors. 

To our knowledge, besides the paper \cite{vuk05}, 
the dynamical system analysis have been used and 
critical points derived 
and analysed mainly for models described by an action of the kind
\beq
S=S_m+\frac{1}{2\chi}\int d^4x\,\sqrt{-g}\,F(R)\,,
\hs\hs
\chi=8\pi G_N\,, 
\label{genR}
\eeq
$F(R)$ being an arbitrary function of the scalar curvature $R$,
$G_N$ the Newton's constant and $S_m$ the matter action.
The simplest choice $F(R)=R-2\Lambda$ corresponds to the $\Lambda$CDM model.
Here we shall propose a method which will permit to derive the critical
points also for models described by a function $F(U_k)$, $U_k$ being
generic geometric (higher-order) invariants (see the Appendix). 
 
Their interest has been 
recently triggered by the appearance of Refs.~\cite{turner,cappozziello} 
and after that investigated in many papers (see, for example
 \cite{fRbiblio}). Furthermore, very recently in 
\cite{hu07,staro07,odin07,AB,seba},
 new ``viable'' models have been introduced and discussed. 

Also the stability of the solutions has been 
discussed in several places, an incomplete list being 
\cite{cognola05,faraoni,cognola,NO,capo,NO1,NO2,bazeia07,rador,soko,b} and 
earlier references quoted in \cite{staro07}. 
To this aim, different techniques have been employed, 
including manifestly covariant and field theoretical 
approaches, where the gauge issue has been properly taken into account. 
All these investigations are in agreement with the following conditions
which ensures the existence and the stability of the de Sitter solution
with scalar curvature $R=R_0$:
\beq
\ag 
\begin{array}{ll}
2F(R_0)-R_0F'(R_0)=0\,,& \mbox{existence,}
\\
\frac{F'(R_0)}{R_0F''(R_0)}-1>0\,,&\mbox{stability,}
\end{array}\cp
\label{R-CP-ST}
\eeq
where $F'$ and $F''$ are the derivatives of $F(R)$ with respect to $R$
and everywhere in the paper the subscript $0$ indicates quantities evaluated 
at the critical point.
The first condition in (\ref{R-CP-ST}) determines the scalar curvature of the 
de Sitter solution, while the second one
gives the condition for the stability around such a solution. 

There are some theoretical (quantum effects and string-inspired) 
motivations in order to investigate 
gravitational models depending on higher-order invariants.
The ``string-inspired'' scalar-Gauss-Bonnet gravity case $F(R,G)$ 
has been suggested in Ref.~\cite{sasaki} as a model for
gravitational dark energy, while some time ago it
has been proposed as a possible solution of the initial singularity 
problem \cite{ART}. 
The investigation of different
regimes of cosmic acceleration in such gravity models
has been carried out in 
Refs.~\cite{sasaki,fGB,Sami,Mota,Calcagni,Neupane,GB,cognola06,cognola066,sami,E}.
In particular, in \cite{cognola06} a first attempt to the study of 
the stability of such kind of models has been carried out
using an approach based on quantum field theory. 
 
The method we shall use in the present paper is based on a 
classical Lagrangian formalism \cite{vilenkin85,capozziello02,cognolaV}, 
inspired by the seminal paper \cite{staro}, 
where quantum gravitational effects were considered for the first time.
With regard to this, it is well known that one-loop and two-loops quantum 
effects induce higher derivative gravitational terms in the effective 
gravitational Lagrangian. Instability due to 
quadratic terms have been investigated in \cite{staro1}. 
A particular case has been recently studied in \cite{aco} 
and general models depending on quadratic invariants
have been investigated in \cite{herv,topo}. 

A stability analysis of nontrivial vacua in a general class of 
higher-derivative theories of 
gravitation has already been presented in \cite{waldram}. 
Our approach is different from the one presented there 
since we are dealing with scalar quantities and moreover it is 
more general, since it is not restricted to the vacuum invariant submanifold.

Finally, it should be stressed that the stability studied here is the one
with respect to homogeneous perturbations. For the $F(R)$ case, the stability
criterion for homogeneous perturbations is equivalent to the inhomogeneous one
\cite{faraoni}. 

The content of the paper is the following. 
In Section 2 we describe our general method in the vacuum case and
derive general conditions for existence and stability of de Sitter 
solutions for arbitrary $F(U_k)$ models. 
Then, in Section 3, we extend the method in order to include 
matter/radiation and give the conditions which (in principle)
permit to derive all critical points.
In Section 4 we analyse some specific models and finally we
make some conclusions in Section 5.
For the reader convenience, 
the paper ends with an Appendix in which we derive some 
interesting useful relations. 

\s{The general case without matter}
\label{S:general}

In this Section we would like to study the stability condition of de Sitter
solutions for models
in which the Lagrangian density is an arbitrary function
of all algebraic invariants built up with the Riemann 
tensor of the FRW space-time we are dealing with, that is
\beq 
{\cal L}=-\frac{1}{2\chi}\,F(R,P,Q,...)\,,
 \eeq
where $R$ is the scalar curvature, $P=R^{\mu\nu}R_{\mu\nu}$ and 
$Q=R^{\alpha\beta\gamma\delta}R_{\alpha\beta\gamma\delta}$ are the two 
quadratic invariants 
and the dots means other independent algebraic invariants of higher order. 

For the sake of convenience we write the metric in the form
\beq 
ds^2=-e^{2n(t)}dt^2+e^{2\al(t)}d{\vec x\,}^2\,,
\hs N(t)=e^{n(t)}\,,\qquad a(t)=e^{\al(t)}\,.
 \eeq 
In this way $\dot\al(t)=H(t)$ is the Hubble parameter and
the generic invariant quantity $U$ has the form
\beq 
U=e^{-2pn(t)}\,u(\dot n,\dot\al,\ddot\al)=e^{-2pn(t)}\,u(\dot n,H,\dot H)
=H^{2p}\,e^{-2pn(t)}\,u(X)\,,
\label{INV}\eeq
where $2p$ is the dimension (in mass) of the invariant under consideration
and $X=(\dot H/H^2-\dot n/H)$ (see the Appendix).
In particular one has
\beq\begin{array}{l}
R=6e^{-2n}\aq 2{H}^{2}+\dot{H}-\dot{n}\,{H}\cq
=6H^2e^{-2n}(2+X)\,,
\\ 
P=12e^{-4n}\left(
\dot{n}^{2}{H}^2
 -3\dot{n}{H}^{3}
 -2\dot{n}{H}\dot{H}
 +3{H}^{4}
 +3{H}^{2}\dot{H}
 +\dot{H}^{2}\right)
=12H^4e^{-4n}(3+3X+X^2)\,,
\\ 
Q=12e^{-4n}\left(
\dot{n}^{2}{H}^{2}
 -2\dot{n}{H}^{3}
 -2\dot{n}{H}\dot{H}
 +2{H}^{4}
 +2{H}^{2}\dot{H}
 +\dot{H}^{2}
\right)
=12H^4e^{-4n}(2+2X+X^2)\,,
\\ 
G=R^2-4P+Q=24e^{-4n}(H^4+H^2\dot H)=24H^4e^{-4n}(1+X)\,.
\end{array}
\label{INV-Q}\eeq
Using this notation, the action reads
\beq 
S=-\int\,d^3x\int\,dt\, L(n,\dot n,\al,\dot\al,\ddot\al)
 =\frac{1}{2\chi}\,\int\,d^3x\int\,dt\,e^{n+3\al}\,F(n,\dot n,\dot\al,\ddot\al)
\label{GrA}\eeq
and the Lagrange equations corresponding to the two Lagrangian variables 
$n(t)$ and $\al(t)$ are given by
\beq 
E_n=\frac{\partial L}{\partial n}
 -\frac{d}{dt}\,\frac{\partial L}{\partial\dot n}
 =0\,,
\label{En}\eeq
\beq 
E_{\al}=\frac{\partial L}{\partial \al}
 -\frac{d}{dt}\,\frac{\partial L}{\partial\dot\al}
 +\frac{d^2}{dt^2}\,\frac{\partial L}{\partial\ddot\al}
 =0\,.
\label{Ea}\eeq
Since the function $n(t)$ is a ``gauge function'', which corresponds to the
choice of the evolution parameter, we expect equation
(\ref{Ea}) to be a direct consequence of 
equation (\ref{En}) and in fact a straightforward calculation 
leads to the identity (see Eq.~(\ref{dEnEa}) in the Appendix)
\beq 
\frac{dE_n}{dt}=\dot nE_n+\dot\al E_{\al}\,.
\label{identita}\eeq
As a consequence of such an identity
we may limit our analysis to the use of 
equation (\ref{En}), much simpler than equation (\ref{Ea}).
Furthermore, we may use the gauge freedom and fix the cosmological 
time by means of the condition $N(t)=1$, that is $n(t)=0$. 
From now on it is understood that all quantities will be evaluated in 
such a gauge and so the parameter $t$ corresponds to the standard cosmological time. 
In this way Eqs.~(\ref{En}) and (\ref{Ea}) read
\beq 
F+F_n-3HF_{\dot n}-\dot F_{\dot n}=H\dot F_{\dot H}-HF_H
+F-\dot HF_{\dot H}+3H^2F_{\dot H}=0\,,
\label{En2}\eeq
\beq 
\ddot F_{\dot H}-\dot F_H+6H\dot F_{\dot H}-3HF_H
+3F+3\dot HF_{\dot H}+9H^2F_{\dot H}=0\,,
\label{Ea2}\eeq
where $F_X=\partial_X F$ means derivative of $F$ with respect to $X$.
All derivatives with respect to $n$ in 
Eq.~(\ref{En2}) have been eliminated by using 
Eq.~(\ref{FH}) in the Appendix.

In the next Section we shall study all critical points of the system 
in the presence of matter, while here we shall limit ourselves
to determine the conditions for the
existence and the stability of de Sitter solutions in the absence of
matter. 
To this aim we take into account Eq.~(\ref{En2}) and re-write 
it as a differential equation for the Hubble parameter $H(t)$, that is
\beq 
HF_{\dot H\dot H}\,\ddot H+H(F_{H\dot H}-F_{\dot H})\,\dot H+\Si(H,\dot H)=0\,,
\label{EMotion}\eeq
where we have set
\beq
\Si(H,\dot H)=F-HF_H+3H^2F_{\dot H}\,.
\label{Sigma}\eeq
We see that the condition for the existence of a de Sitter solution 
$P_0\equiv(\dot H=0,H=H_0)$ is
\beq 
\Si(H_0,0)=0 \segue \aq F-HF_H+3H^2F_{\dot H}\cq_{P_0}=0\,,
 \eeq
where here all quantities have to be evaluated at $P_0$.
The latter condition, which gives rise to the de Sitter critical point,
can be simplified by means of Eq.~(\ref{FhFh})
in the Appendix. In fact we have
\beq 
\aq F-HF_H+3H^2F_{\dot H}\cq_{P_0}=\aq F-H^2F_{\dot H}\cq_{P_0}=0\,.
\label{stab1}\eeq
In order to study the stability of such a solution 
we have to distinguish between two
possible cases. The simplest one occurs when
$F$ is linear in $\dot H$, that is 
$F_{\dot H\dot H}=0$. 
In such a case Eq.~(\ref{EMotion}) assumes the form
\beq 
(HF_{H\dot H}-HF_{\dot H}-\Si_{\dot H})\dot H+\Si(H,0)=
\Si(H,0)=0\segue H=H_0\,.
\label{stab2}
\eeq
Then we see that in such a special case the solution,
if it exists, is trivially stable, since 
the field equation reduces
to an algebraic equation which fixes the value of
$H$ in terms of the parameters of the system. 
The Einstein's theory with cosmological constant $F=R-2\La$ is
an example of this kind and in fact Eq.~(\ref{stab2}) gives rise to 
$H^2=H_0^2=\La/3$.

In the second case $F_{\dot H\dot H}\neq0$,
in order to study the stability one may transform Eq.~(\ref{EMotion}) in
an autonomous system by introducing the functions $K(t)=\dot H(t)$. 
In this way, one gets
\beq 
\dot H=K\,,\hs\hs
\dot K=-\frac{\Si(H,K)+(HF_{H\dot H}-F_{\dot H})\,K}{HF_{\dot H\dot H}}\,.
 \eeq
The critical points are determined by the condition $(\dot H=0,\dot K=0)$ 
and the linearized system which determines the stability reads
\beq
\at\begin{matrix}{\de \dot H\cr\de\dot K\cr}\end{matrix}\ct
=M\,\at\begin{matrix}{\de H\cr\de K\cr}\end{matrix}\ct
\hs\hs
M=\at\begin{matrix}{0&1\cr A_0&B_0\cr}\end{matrix}\ct\,,
\eeq
where
\beq
A_0&=&\aq-\frac{\Si_H}{HF_{\dot H\dot H}}\cq_{P_0}
=\aq\frac{F_{HH}-3HF_{H\dot H}-6F_{\dot H}}{F_{\dot H\dot H}}\cq_{P_0}\,,
\eeq
\beq
B_0&=&\aq\frac{F_{\dot H}-HF_{H\dot H}-\Si_{\dot H}}{HF_{\dot H\dot H}}\cq_{P_0}
=-H_0\,.
\eeq
The stability conditions are obtained by requiring 
$\Tr M<0$ and $\det M>0$. The first condition is trivially satisfied
since $H_0$ is positive,
while the second one gives
\beq 
\aq\frac{3HF_{H\dot H}-F_{HH}+6F_{\dot H}}{F_{\dot H\dot H}}\cq_{P_0}>0\,,
\label{stabilita}\eeq

Summarizing, we have the general results
\beq 
\ag\begin{array}{ll}  
\aq F-H^2F_{\dot H}\cq_{P_0}=0\,,&\mbox{critical points,}\\
\aq\frac{3HF_{H\dot H}-F_{HH}+6F_{\dot H}}{F_{\dot H\dot H}}\cq_{P_0}>0\,,
&\mbox{stability condition when }F_{\dot H\dot H}\neq0\,,
\end{array}\cp
\label{CP-ST}\eeq
In the particular case in which $F_{\dot H\dot H}=0$ the eventual
(de Sitter) critical point is always stable. 
Of course, when $F=F(R)$, Eqs.~(\ref{CP-ST}) become equivalent to Eqs.~(\ref{R-CP-ST}).

\section{The general case with matter}
\label{S:matter}
The inclusion of matter in the model is obtained by adding 
the matter action $S_m$ to the gravitational action (\ref{GrA}).
Of course, in order to preserve symmetry we have to consider 
homogeneous and isotropic matter, that is a perfect fluid.
First of all we observe that 
\beq 
\frac{\de S_m}{\de n}\,\de n&=&-\int\,d^4x\sqrt{-g}\,T_{00}g^{00}\,\de n\,,
\\
\frac{\de S_m}{\de\al}\,\de\al&=&-\int\,d^4x\sqrt{-g}\,T_{ab}g^{ab}\,\de\al\,,
 \eeq
where $T_{\mu\nu}\equiv(T_{00},T_{ab})$ is the energy-momentum tensor, which
for a perfect fluid satisfies the conditions
\beq 
T_{00}g^{00}=-\rho\,,\hs\hs T_{ab}g^{ab}=3p\,,\hs\hs a,b=1,2,3\,,
\nn \eeq
$\rho$ and $p=p(\rho)$ being respectively the density and the pressure
of matter.

Now Eqs.~(\ref{En}) and (\ref{Ea}) trivially changes for the presence 
of matter and in fact we get
\beq 
E_n=\frac{\partial L}{\partial n}
 -\frac{d}{dt}\,\frac{\partial L}{\partial\dot n}
 =2\sqrt{-g}\,T_{00}g^{00}=2\rho\,\sqrt{-g}
\,,
\label{EnM}\eeq
\beq 
E_{\al}=\frac{\partial L}{\partial \al}
 -\frac{d}{dt}\,\frac{\partial L}{\partial\dot\al}
 +\frac{d^2}{dt^2}\,\frac{\partial L}{\partial\ddot\al}
 =\sqrt{-g}\,T_{ab}g^{ab}=-6p\,\sqrt{-g}\,.
\label{EaM}\eeq
The identity (\ref{identita}) is still valid and is equivalent to the
energy-momentum conservation law
\beq 
\nabla_\mu T^{\mu\nu}=0\segue \dot\rho=-3H\,(\rho+p)\,.
\label{cons}\eeq
Using the notation of the previous Section and putting again $n(t)=0$
now we get
\beq 
H\dot F_{\dot H}-HF_H+F-\dot HF_{\dot H}+3H^2F_{\dot H}=2\rho\,,
\label{EnM2}\eeq
\beq 
\ddot F_{\dot H}-\dot F_H+6H\dot F_{\dot H}-3HF_H
+3F+3\dot HF_{\dot H}+9H^2F_{\dot H}=-6p\,.
\label{EaM2}\eeq
The latter equations are the generalisation to arbitrary action of
the well known Friedmann equations. It is interesting to observe that in the
pure Einstein gravity, that is for $F=R$,
they read
\beq
H^2F_{\dot H}=F_X=2\rho\segue \Om_\rho=1\,,
\label{EnS}\eeq
\beq
(3H^2+2\dot H)F_{\dot H}=(3+2X)F_X=-6p
\segue \Om_p=-1-\frac23\,X\,,
\label{EaS}\eeq
where we have introduced the dimensionless variables
\beq
\Om_\rho=\frac{2\rho}{H^2F_{\dot H}}=\frac{2\rho}{F_X}\,,\hs\hs
\Om_p=\frac{2p}{H^2F_{\dot H}}=\frac{2p}{F_X}\,,
\label{Om-rp}\eeq
which in this special case are given by the usual values
$\Om_\rho=\rho/3H^2$ and $\Om_p=p/3H^2$.
From Eqs.~(\ref{EnS}) and (\ref{EaS}) it follows
\beq
w\equiv\frac{p}{\rho}=\frac{\Om_\rho}{\Om_p}=-1-\frac23\,X\,.
\label{w}\eeq
In the general case, Eqs.~(\ref{EnM2}) and (\ref{EaM2}) have
more terms with respect to (\ref{EnS}) and (\ref{EaS})
and it is quite natural to interpret them as corrections
due to the presence of higher-order terms in the action.
Then we define
\beq
\Om_{\rho}^{eff}=\Om_\rho+\Om_\rho^{curv}=1\,,\hs\hs
\Om_p^{eff}=\Om_p+\Om_p^{curv}=
-1-\frac23\,X\,,
\eeq
\beq
w_{eff}\equiv\frac{\Om_p^{eff}}{\Om_\rho^{eff}}=-1-\frac23\,X\,,
\label{weff}\eeq
where $\Om_\rho^{curv}$ and $\Om_p^{curv}$ are complicated expressions,
which only depend on the function $F$. They can be derived
from (\ref{EnM2}) and (\ref{EaM2}),
but they explicit form is not necessary for our aims.
The effective quantity $w_{eff}$ is equal to the ratio between the
effective density and the effective pressure and it could be
negative even if one considers only ordinary matter.
It is known that the current-measured value of $w_{eff}$
is near $-1$.

In order to get all critical points of the system now we follow the
method described in \cite{amendola}. First of all we introduce the
dimensionless variables
\beq
\Om_\rho=\frac{2\rho}{H^2F_{\dot H}}=\frac{2\rho}{F_X}\,,\hs\hs
\Om_p=\frac{2p}{H^2F_{\dot H}}=\frac{2p}{F_X}\,,
 \eeq
\beq
X=\frac{\dot H}{H^2}\,,\hs
Y=\frac{F-HF_H}{H^2F_{\dot H}}=\frac{F}{F_X}-X\,,\hs
Z=\frac{\dot F_{\dot H}-F_H}{HF_{\dot H}}=\frac{F'_X}{F_X}-2X-\xi\,,
\label{XYZ}\eeq
where the prime means derivative with respect to $\al$ and the quantity
\beq 
\xi=\xi(X,Y)=\frac{F_H}{HF_{\dot H}}=\frac{HF_H}{F_X}\,,
\label{xi}\eeq
has to be considered as a function of the variables $X$ and $Y$. 
In general it is a function of $X$ and $H$, but this latter 
quantity can be expressed in terms of $X$ and $Y$ as a direct 
consequence of the definition of $Y$ itself.
Then we derive an autonomous system by taking the 
derivatives of such variables.
From Eq.~(\ref{EnM2}) we have the constraint
\beq 
\Om_\rho=Y+Z+3\segue\Om_{curv}=-(Y+Z+2)\,,
\label{vincolo}\eeq
which reduces to the standard one when $F$ is linear in $R$ (general relativity with
cosmological constant). 

Deriving the variables above by taking into account of 
(\ref{cons}) and (\ref{EaM2}) we get the system
of first order differential equations 
\beq 
\ag\begin{array}{l}
X'=-2X^2-\ga X+\be (Z+\xi)\,,
\\ 
Y'=-(2X+Z+\xi)Y-XZ\,,
\\ 
Z'=-3\Om_p-[X(Z+6)+Z(Z+\xi+6)+3(Y+\xi+3)]\,,
\\ 
\Om'_\rho=-3(\Om_\rho+\Om_p)-(2X+Z+\xi)\Om_\rho\,,
\\ 
\Om'_p=-3\frac{dp}{d\rho}\,(\Om_\rho+\Om_p)-(2X+Z+\xi)\Om_p\,,
\end{array}\cp
\label{AutSys}\eeq
where $X'\equiv\frac{dX}{d\al}=\frac{1}{H}\frac{dX}{dt}$ and so no and 
$p=p(\rho)$ has been assumed.
The first equation in (\ref{AutSys}) has been obtained by deriving the function $F_{\dot H}$
with respect to $t$ and putting
\beq 
\be=\be (X,Y)=\frac{F_{\dot H}}{H^2F_{\dot H\dot H}}
 =\frac{F_X}{F_{XX}}\,,\hs\hs
\ga=\ga(X,Y)=\frac{F_{H\dot H}}{HF_{\dot H\dot H}}
 =\frac{HF_{HX}}{F_{XX}}=\be\xi_X+\xi\,.
\label{bega}\eeq
It is understood that $F_{\dot H\dot H}\neq0$ has been assumed.

Note that in principle one could consider a mixture 
of different kind of matter/radiation with densities $\rho_k$ and 
corresponding pressures $p_k$. In such a case in Eqs.~(\ref{AutSys})
there will appear the sums of the corresponding quantities
$\Om_{\rho_k}$ and $\Om_{p_k}$ and the corresponding 
differential equations for any type
of matter/radiation. For simplicity here we consider only one type of
matter/radiation described by the equation of state $p=p(\rho)$. 

In general, among the five differential equations (\ref{AutSys})
only three of them are linear independent for the presence
of the constraint (\ref{vincolo}) and the equation of state. 
For simplicity now we assume the pressure to be proportional to the 
density, that is $p=w\rho$, with constant $w$, where for ordinary 
matter $0\leq w\leq1/3$
($w=0$ corresponds to dust, while $w=1/3$ to pure radiation),
but in principle one could also consider ``exotic'' matter with $w<0$
and cosmological constant which corresponds to $w=-1$. 
With this choice and from Eq.~(\ref{vincolo}) we get
\beq 
\Om_p=w\Om_\rho\,,\hs\hs \Om_\rho=Z+Y+3\,.
 \eeq
As a consequence the autonomous system which gives rise to the critical points
can be chosen as
\beq 
\ag\begin{array}{l} 
0=X'=-2X^2-\ga X+\be(Z+\xi)
\\
0=Y'=-(2X+Z+\xi)Y-XZ
\\
0=Z'=-3(1+w)(Z+Y+3)-(Z+\xi)(Z+3)-X(Z+6)
\end{array}\cp
\label{AutS}\eeq
and $\Om_\rho$ at the critical points will be determined by means 
of Eq.~(\ref{vincolo}).
The critical points are the solutions of the algebraic 
system (\ref{AutS}). 
The number and the position of such points depends on the Lagrangian throughout
the functions $\be,\ga$ and $\xi$. 
In principle, given $F$ one can derive 
all critical points as the constant solutions of Eqs.~(\ref{AutS}),
but in practice for a generic $F$ the algebraic system could 
be very complicated and the solutions quite involved.
We shall consider in detail some particular cases in the next Section.

A brief comment about the equivalence of (\ref{AutS}) and 
Eqs.~(\ref{EnM2})-(\ref{EaM2}) is in order. It is well known that
the Gauss-Bonnet scalar $G$ does not contribute to the field equations,
because it is a topological invariant and this means that the field
equations (\ref{EnM2})-(\ref{EaM2}) are invariant with respect to
the transformation $F\to F+c\,G$, $c$ being an arbitrary constant. 
On the contrary, the new variables $X,Y,Z$ (by definitions) 
are not invariant with respect to such a transformation and
so the system (\ref{AutS}) could contain the parameter $\ga$ explicitly,
but nevertheless the solutions for $H(t)$ do not depend on such a
parameter. The same thing is true for the critical points.

It has also to be noted that in general, given a Lagrangian $F$,
there are solutions of the system (\ref{AutS}), which
give rise to trivial or unphysical models. Of course such
solutions have to be dropped (see specific examples).

Before to analyse all possible solutions of (\ref{AutS}), 
we observe that in the absence of matter 
the de Sitter solution $X=0$ of course is a critical point for the system
if the first equation in (\ref{CP-ST}) is satisfied. In fact we trivially see that
$X=0$, $\Om_\rho=Y+Z+3=0$ satisfy (\ref{AutS})
if $Y-\xi+3=0$, which is equivalent to first equation in  (\ref{CP-ST}) when $X=0$.
(Strictly speaking, the condition $Y=1$ is equivalent to (\ref{CP-ST}),
when $H_0\neq0$. In principle, Eq.~(\ref{CP-ST}) can also have the
Minkowskian solution $R_0=0$). 
On the de Sitter solution the value of $w_{eff}$ is exactly equal to
$-1$. 
By taking the variations of (\ref{AutS}) with respect to $X$ and $Y$ 
on the critical point and using the identity (\ref{FhFh})
in the Appendix we also get the stability condition in (\ref{CP-ST}).

In order to show that the present approach is a generalisation of
the one proposed in \cite{amendola}, 
we observe that when the function $F=F(R)$ depends
on the unique invariant $U=R$, we exactly obtain 
all known results.
For such a special case, using (\ref{INV-Q}) (with $n(t)=0$)
and (\ref{UX})-(\ref{FH}) in the appendix we get
\beq 
\xi&=&\frac{HF_H}{F_X}=\frac{2pU}{U_X}-2X=4\,,\\
\be &=&\frac{F_X}{F_{XX}}=\frac{F_UU_X}{F_UU_{XX}+F_{UU}U_X^2}
=\frac{F_R}{6H^2F_{RR}}\,,\\
\ga &=&\frac{HF_{HX}}{F_{XX}}=\be\xi_X+\xi=4\,,
 \eeq
Now with the trivial replacement
\beq 
X=x_3-2\,,\hs Y=x_2-x_3+2\,,\hs Z=-(x_1+4)\,,
\hs\hs\be=\frac{x_3}{m}\,,
 \eeq
Eqs.~(\ref{AutSys}) become equivalent to the analog equations in 
Ref.~\cite{amendola}, where this class of models have been studied in detail.

As already said above, the critical points of (\ref{AutS}) 
depends on $\xi,\be,\ga$, which in general are complicated 
functions of $X$ and $Y$, then it is not possible to determine
general solutions without to choose the model, nevertheless it is
convenient to distinguish two distinct classes of solutions
characterised by the values of $w\neq-1$ and $w=-1$.
For the sake of completeness we consider $w\leq1/3$ and so 
we write the solutions also for ``exotic'' matter, that is
quintessence ($-1<w<0$) and phantom ($w<-1$). 
Of course, such solutions have to be dropped if
one is only interested in ordinary matter/radiation.
We have
\begin{itemize}
\item $w\neq-1$ --- The critical points
are the solutions of the system of three equations
\beq 
\ag\begin{array}{l}
2X^2+\ga X-\be(Z+\xi)=0
\\
(2X+Z+\xi)Y+XZ=0
\\
3(1+w)(Z+Y+3)+(Z+\xi)(Z+3)+X(Z+6)=0
\end{array}\cp\hs\hs w\neq-1\,,
\label{C1}\eeq
where $\xi,\be,\ga$ are functions of $X,Y$ determined by Eqs.~(\ref{xi})
and (\ref{bega}). 
The stability matrix has three eigenvalues and the point is 
stable if the real parts of all of them are negative. 

The latter system has always the de Sitter
solution $P_0\equiv(X=0,Y=1,Z=-4)$, where $\Om_\rho=0$
and $w_{eff}=-1$. 
Note however that such a solution could exist also in the presence
of matter, since the existence of $P_0$ critical point 
only implies that the critical value for $\Om_\rho$ vanishes.

\item $\Om_\rho\neq0$, $w=-1$ --- 
The critical points are given by
\beq 
\ag\begin{array}{l}
2X^2+\ga X-\be(Z+\xi)=0
\\
(2X+Z+\xi)Y+XZ=0
\\
(Z+\xi)(Z+3)+X(Z+6)=0
\end{array}\cp
\hs\hs w=-1\,.
\label{C-1}\eeq
For this class of solutions, the non-singular stability matrix
has three eigenvalues and the point is stable if the real parts of all of
them are negative.

We see that there is at least one singular case 
(critical line) when $X=0$ and $Z=-\xi=-4$. 
In fact in such a case $Y$ or $\Om_\rho$ are undetermined since
\beq
\Om_\rho=Y+3-\xi(0,Y)=Y-1\segue Y=1+\Om_\rho\,,
\hs\hs\Om_\rho\mbox{ arbitrary}\,.
\label{special}\eeq
Such a solution can be seen as a generalisation of 
the de Sitter solution for a model with cosmological constant.
The de Sitter critical point for the model $\tilde F=F-2\La$
reads $(X=0,\tilde Y=1,\tilde Z=-4)$.
Such a solution follows from Eq.~(\ref{special}) if we choose
$\rho_0\equiv\La$. In fact,
on the critical point $(X=0,Y=1+\Om_\rho,Z=-4)$ (Eq.~(\ref{special})) 
and from definitions (\ref{XYZ}) we get
\beq 
\Om_\rho=\frac{2\rho}{H^2F_{\dot H}}=\frac{F}{H^2F_{\dot H}}-1
\segue\tilde Y\equiv\frac{\tilde F}{H^2\tilde F_{\dot H}}=1\,,
 \eeq
which corresponds to de Sitter critical point for $\tilde F$.
Of course, Eq.~(\ref{special}) is more general than the
case with pure cosmological constant since $\rho$ is not
necessary a constant. 

Of course for this special class of solutions $w_{eff}=-1$. 
Note also that the stability matrix has always a vanishing eigenvalue
and the stability of the system is determined by the
other two eigenvalues.

For some models, but just for technical reasons, 
it could be convenient to treat the cosmological constant as matter,
using the previous identification we have done. 

\end{itemize}

\section{Explicit examples}

In order to see how the method works, 
now we give explicit solutions for some models and, when possible we also 
study the stability of the critical points. 
We restrict our analysis to the values $0\leq w\leq1/3$ and
to the special value $w=-1$, which corresponds to the pure cosmological constant,
but in principle any negative value of $w$ could be considered,
even if this will be in contrast with the aim of modified gravity.
In fact, modified gravity can generate an effective negative value of $w$ 
without the use of phantom or quintessence.

It as to be stressed that in general, due to technical difficulties,
one has to study the models by a numerical analysis. 
Only for some special cases one is able to
find analytical results. 
Here we report the results
for some models of the latter class in which the analytical 
analysis can be completely carried out. We also study
more complicated models and for those we limit
our analysis to the de Sitter solutions. 

In the following we shall use the compact notation
\beq 
P\equiv(X,Y,Z,\Om_\rho,w_{eff})\,,\hs
P_0\equiv(0,1,-4,0,-1)\,,\hs
P_\La=(0,1+\Om_\La,-4,\Om_\La,-1)\,.
 \eeq
The latter is an additional critical point
that we have for the choice $w=-1$ and can be seen as the de Sitter solution
in the presence of cosmological constant.

\begin{description}
\item
$F=R-\mu^4/R$ --- This is the well known model introduced in 
\cite{turner,cappozziello} and discussed in \cite{amendola}. 
For this model the system (\ref{AutS}) with arbitrary $w$ 
has six different solutions, 
but only two of them effectively correspond to 
physical critical points, if $0\leq w\leq1/3$. 
In principle there are other critical points for negative
values of $w$ (phantom or quintessence) and 
moreover there is also a particular solution for $w=-1$ which 
corresponds to the model with a cosmological constant $\La$. 
 
Solving the autonomous system one finds
\begin{itemize}
\item 
$P=P_0$: unstable de Sitter critical point.
The critical value for the scalar curvature reads $R_0=\sqrt3\,\mu^2$.
\item 
$P=(-1/2,-1,-2,0,-2/3)$: stable 
critical point. At the critical value, $H_0=0$. 
\item 
$P=(3(1+w)/2,-(5+3w),-2(5+3w),-3(4+3w),-(2+w))$: 
unstable critical point where $H_0=0$. 
\item 
$P=P_\La$: unstable critical point. At the critical value one has 
$
H_0^2=(\La/6)(1+\sqrt{1+3\mu^4/4\La^2}\,.
$
\end{itemize}

\item
$F=R+aR^2+bP+cQ$ --- (Starobinsky-like model).
Here we have to assume $3a+b+c\neq0$ otherwise the quadratic term
becomes proportional to the Gauss-Bonnet invariant. 
For $0\leq w\leq1/3$, this model has only one critical point. 
In order to have a de Sitter solution, we have to introduce a 
cosmological constant $\La$. We have in fact
\begin{itemize}
\item 
$P=P_0$: Minkowskian solution 
with $R_0=0$, which is stable if $3a+b+c>0$.
\item 
$P=P_\La$:
de Sitter critical point with $R_0=6\La$, which 
is stable if $3a+b+c>0$, in agreement with \cite{topo}.
\end{itemize}

\item
$F=R-d^2Q_3$ --- This is the simplest toy model with 
the cubic invariant
$Q_3=R^{\al\be\ga\de}R_{\al\be\mu\nu}R^{\mu\nu}_{\ga\de}$.
For this model we have the following critical points:
\begin{itemize}
\item
$P=P_0$: unstable de Sitter solution with $R_0=6/d$.
\item
$P=P_0$: stable Minkowskian solution with $R_0=0$.
\item
$P=(0.05,0.60,-3.60,0,-1.03)$: stable solution with $H_0=0$.
\item
$P=P_\La$: this point exists, but not for any value of $d$ and $\La$.
Also the value of $H_0$ and the stability depend on the parameters.
 \end{itemize}

\item
$F=R+aR^2+bP+cQ-d^2Q_3$ --- This is a generalisation of the previous
two models. It may be motivated by the two-loop corrections in quantum gravity 
\cite{sagno,van}. The de Sitter critical points have been studied in 
Ref.~\cite{monica}.
The algebraic equations (\ref{AutS}) are too 
complicated to be solved analytically, but it is easy
to verify that there are at least the following solutions:
\begin{itemize}
\item
$P=P_0$: de Sitter solution with $R_0=6/d$. This is stable 
if $3a+b+c+3d>0$.
\item 
$P=P_0$: Minkowskian solution 
with $R_0=0$, which is stable if $3a+b+c>0$.
\item
$P=P_\La$: also in this case this point exists and 
is stable depending on the parameters (see \cite{monica}).

 \end{itemize}

\end{description} 

\s{Conclusion}
In this paper we have presented a general technique which permits 
to arrive at a first order autonomous system of differential equations 
classically equivalent to the equations of motion for models of 
modified gravity based on an arbitrary function $F(R,P,Q,Q_3...)$, 
namely built up with all possible geometric invariant quantities of 
the FRW space-time. Dynamical system techniques have been applied to the investigation of critical points. We have shown that, in the special case
of $F(R)$ theories, the method gives rise 
to the well known results \cite{amendola,amendola1,dunsby}, 
but in principle it can be applied to the study of 
much more general cases. 

As applications, we have considered some
simple models, for which a complete analytical 
analysis have been carried out. However, in general,
due to technical difficulties, a numerical analysis 
is required. Among the models investigated, we would like to remind that 
we were able to deal with one which involves a cubic invariant in the 
curvature tensor and, to our knowledge, this has never been considered before,
and this shows the power of our approach. 

\appendix
\section{Appendix}
Here we show that in the cases we are considering, Eq.~(\ref{identita})
is always satisfied. 
To this aim, as in Eq.~(\ref{INV}), we denote by $U(x)$
a generic invariant quantity of dimension $2p$ (in mass) computed in the
system of coordinates $\{x\}\equiv\{(t,\vec x)\}$ and by
$\tilde U(y)$ the same quantity computed in the system
$\{y\}\equiv\{(\tau,\vec y)\}$. Such coordinates are
chosen in such a way that $d\tau=e^{n(t)}dt$ and as a consequence
the metrics read
\beq 
ds^2=-e^{2n(t)}dt^2+e^{2\al(t)}d{\vec x\,}^2=
 -d\tau^2+e^{2\tilde\al(\tau)}d{\vec y\,}^2\,.
 \eeq
Recalling that $U$ is built up with the Riemann tensor, 
for dimensional reasons in the system $\{y\}$ one has
\beq 
\tilde U(y)=\sum_{k=0}^p\,
 A_{k}\aq\frac{d\tilde\al(\tau)}{d\tau}\cq^{2(p-k)}\,
 \aq\frac{d^2\tilde\al(\tau)}{d\tau^2}\cq^k\,,
 \eeq
and since under diffeomorphism $U$ is a scalar quantity we also have
\beq 
U(x)=\tilde U(y)=e^{-2p n(t)}\,\sum_{k=0}^p\,
 A_{k}\aq\frac{d\al(t)}{dt}\cq^{2(p-k)}\,
 \aq\frac{d^2\al(t)}{dt^2}-\frac{d\al(t)}{dt}\,\frac{dn(t)}{dt}\cq^k\,.
 \eeq
This means that in a generic system of coordinates an arbitrary invariant
can be written in the form
\beq 
U=e^{-2pn}\,H^{2p}\,\sum_{k=0}^p\, A_{k}X^k\,,\hs\hs
X=\frac{\ddot\al}{\dot \al^2}-\frac{\dot n}{\dot\al}=
\frac{\dot H}{H^2}-\frac{\dot n}{H}\,.
\label{U}\eeq
From the latter equation we directly get
\beq
\begin{array}{ll} 
U_{\dot n}=-\frac{U_X}{H}\,,&\hs\hs U_n=-2pU\,,
\\
U_{\dot H}=\frac{U_X}{H^2}\,,&\hs\hs
U_H=\frac{2pU}{H}-\frac{2\dot H\,U_X}{H^3}+\frac{\dot n\,U_X}{H^2}
\end{array}
\label{UX}\eeq
and for a generic function $F(U_1,U_2,U_3,...U_a,...)$
\beq 
\begin{array}{ll} 
F_{\dot n}=-\frac{F_X}{H}\,,&\hs\hs F_n=-2\sum_ap_aU_aF_{U_a}\,,
\\
F_{\dot H}=\frac{F_X}{H^2}\,,&\hs\hs 
F_H=\frac2H\sum_ap_aU_aF_{U_a}-\frac{2X\,F_X}{H}-\frac{\dot n\,F_X}{H^2}\,.
\end{array}
\label{FH}\eeq
From equations above it directly follows
\beq 
F_n+HF_H+2\dot HF_{\dot H}+\dot nF_{\dot n}=0\,.
\label{dUF}\eeq
After these considerations it is quite easy to recover Eqs.~(\ref{identita}).
In fact, using Eqs.~(\ref{En}), (\ref{Ea}), (\ref{FH})-(\ref{dUF}) we get
\beq 
\dot E_n-\dot n E_n-\dot\al E_\al&=&\frac{e^{n+3\al}}{2}\,\aq
(\dot n+3\dot\al)(F_n+HF_H+2\dot HF_{\dot H}+\dot nF_n)
\cp\nn\\&&\hs\hs\hs\ap
+\frac{d}{dt}\,(F_n+HF_H+2\dot HF_{\dot H}+\dot nF_n)\cq
=0\,.
\label{dEnEa}\eeq
Before to end the appendix we also derive the useful relation
\beq 
\aq\frac{F_H}{HF_{\dot H}}\cq_{(X=0,n=0)}=4\segue
\aq H\partial_H\log\frac{F_H}{F_{\dot H}}\cq_{(X=0,n=0)}=1\,.
\label{FhFh}\eeq
The latter identity can be derived as follows.
From (\ref{FH}) we have
\beq 
\aq\frac{F_H}{HF_{\dot H}}\cq_{(X=0,n=0)}=
\aq\frac{HF_H}{F_X}\cq_{(X=0,n=0)}=\frac{2\sum_a\,p_aA_0(U_a)H^{2p_a}F_{U_a}}
{\sum_a\,A_1(U_a)H^{2p_a}F_{U_a}}\,,
\eeq
where $A_k(U_a)$ are the coefficients of the invariant $U_a$ as in (\ref{U}).
Then eq.~(\ref{FhFh}) is true if for any invariant $U_a$ one has 
\beq
2A_1(U_a)=p_aA_0(U_a)\,.
\label{A0A1}\eeq
Such a relation is a direct consequence of the form
of the Riemann tensor. In fact, in this case the non vanishing components read
\beq 
R_{0a0a}=e^{2\al}H^2(1+X)\,,\hs\hs R_{abab}=-e^{-2n}e^{4\al}H^2\,,\hs\hs a,b=1,2,3
\nn\eeq 
and since any invariant is built up with Riemann tensor it is easy to see
that the relation (\ref{A0A1}) is always satisfied.


\begin{thebibliography}{99}


\bibitem{padmanabhan}V. Sahni and A.A. Starobinsky, Int. J. Mod. Physics {\bf D 9} 373 (2000); S. M. Carroll, Living Rev. {\bf 4} 1 (2001); P. J. E. Peeble and B. Ratra, Rev. Mod. Physisc {\bf 75} 559 (2003); T.~Padmanabhan, Phys.\ Rept.\ {\bf 380} 235 (2003); AIP Conf.\ Proc.\ {\bf 861} (2006) 179 [arXiv:astro-ph/0603114]. 

\bibitem{CST}E.~Copeland, M.~Sami and S.~Tsujikawa, Int.\ J.\ Mod.\ Phys.\ D {\bf 15} (2006) 1753. 

\bibitem{rev3}S.~Nojiri and S.~D.~Odintsov, Int.\ J.\ Geom.\ Meth.\ Mod.\ Phys.\ {\bf 4} (2007) 115 

\bibitem{monica}G. Cognola, M. Gastaldi and S. Zerbini, {\em On the Stability of a Class of Modified Gravitational Models} to appear in Int. J. Theor. Physics (2008), [arXiv:gr-qc/0701138]. 

\bibitem{ellis}J. Wainwright and G. F. R. Ellis, {\em Dynamical Systems in Cosmology}, Cambridge University Press, Cambridge (1997); M.~Goliath and G.~F.~R.~Ellis, Phys.\ Rev.\ D {\bf 60} (1999) 023502. 

\bibitem{amendola}L. Amendola, R. Gannouji, D. Polarski and S. Tsujikawa, Phys. Rev. {\bf D75}083504 (2007). 

\bibitem{amendola1}L. Amendola and S. Tsujikawa, {\em Phantom crossing, equation-of-state singularities, and local gravity constraints in f(R) models } [ArXiv: 0705.0396]. 

\bibitem{dunsby}S.~Carloni, A.~Troisi and P.~K.~S.~Dunsby, {\em Some remarks on the dynamical systems approach to fourth order gravity} arXiv:0706.0452; N. Goheer, J. A. Leach, Peter K. S. Dunsby, Class. Quantum Grav. {\bf 25} (2008) 035013. 

\bibitem{fay}S.~Fay, S.~Nesseris and L.~Perivolaropoulos, Phys.\ Rev.\ D {\bf 76} (2007) 063504. 

\bibitem{vuk05}S. M. Carroll, A. De Felice, V. Duvvuri, D. A. Easson, M. Trodden and M.S. Turner, Phys. Rev. {\bf D71}063513(2005). 

\bibitem{turner}S. M. Carroll, V. Duvvuri, M. Trodden and M.S. Turner, Phys. Rev. {\bf D70}043528(2004). 

\bibitem{cappozziello}S. Capozziello, S. Carloni, A. Troisi, {\em Quintessence without scalar fields } [ArXive: astro-ph/0303041]; S. Capozziello, V.F. Cardone, A. Troisi Phys. Rev. {\bf D71}043503 (2005). 

\bibitem{fRbiblio}A.~W.~Brookfield, C.~van de Bruck and L.~M.~H.~Hall, Phys.\ Rev.\ D \textbf{74}, 064028 (2006); T.~P.~Sotiriou, Class.\ Quant.\ Grav.\ {\bf 23}, 5117 (2006); S.~Nojiri and S.~D.~Odintsov, Phys.\ Rev.\ D {\bf 74}, 086005 (2006); N.~J.~Poplawski, Phys.\ Rev.\ D {\bf 74}, 084032 (2006); A.~Borowiec, W.~Godlowski and M.~Szydlowski, Phys.\ Rev.\ D \textbf{74}, 043502 (2006); T.~Koivisto, Phys.\ Rev.\ D {\bf 73}, 083517 (2006); A.~de la Cruz-Dombriz and A.~Dobado, Phys.\ Rev.\ D \textbf{74}, 087501 (2006); T.~Multamaki and I.~Vilja, Phys.\ Rev.\ D {\bf 74}, 064022 (2006); T.~P.~Sotiriou, Phys.\ Lett.\ B \textbf{645}, 389 (2007); T.~P.~Sotiriou and S.~Liberati, Annals Phys.\ \textbf{322}, 935 (2007); V.~Faraoni and S.~Nadeau, Phys.\ Rev.\ D \textbf{75}, 023501 (2007); D.~Huterer and E.~V.~Linder, Phys.\ Rev.\ D \textbf{75}, 023519 (2007); S.~Fay, R.~Tavakol and S.~Tsujikawa, Phys.\ Rev.\ D \textbf{75}, 063509 (2007); K.~Kainulainen, J.~Piilonen, V.~Reijonen and D.~Sunhede, Phys.\ Rev.\ D {\bf 76}, 024020 (2007); A.~De Felice and M.~Hindmarsh, JCAP {\bf 0706}, 028 (2007); K.~Uddin, J.~E.~Lidsey and R.~Tavakol, Class.\ Quant.\ Grav.\ {\bf 24}, 3951 (2007); J.~C.~C.~de Souza and V.~Faraoni, Class.\ Quant.\ Grav.\ {\bf 24}, 3637 (2007); M.~S.~Movahed, S.~Baghram and S.~Rahvar, Phys.\ Rev.\ D {\bf 76}, 044008 (2007); E.~O.~Kahya and V.~K.~Onemli, Phys.\ Rev.\ D {\bf 76} (2007) 043512; O.~Bertolami, C.~G.~Bohmer, T.~Harko and F.~S.~N.~Lobo, Phys.\ Rev.\ D {\bf 75} (2007) 104016; M.~Fairbairn and S.~Rydbeck, ``Expansion history and f(R) modified gravity,'' arXiv:astro-ph/0701900; P.~J.~Zhang, Phys.\ Rev.\ D \textbf{73}, 123504 (2006); S.~Bludman, ``What Drives Our Accelerating Universe?'', arXiv:astro-ph/0702085; S.~Capozziello and M.~Francaviglia, ``Extended Theories of Gravity and their Cosmological and Astrophysical Applications,'' arXiv:0706.1146 [astro-ph]; A.~De Felice, P.~Mukherjee and Y.~Wang, ``Observational Bounds on Modified Gravity Models,'' arXiv:0706.1197 [astro-ph]; K.~Bamba, Z.~K.~Guo and N.~Ohta, ``Accelerating Cosmologies in the Einstein-Gauss-Bonnet Theory with Dilaton,'' arXiv:0707.4334 [hep-th]; C.~G.~Boehmer, T.~Harko and F.~Lobo, ``Dark matter as a geometric effect in f(R) gravity,'' arXiv:0709.0046 [gr-qc]. 

\bibitem{hu07}W.~Hu and I.~Sawicki, Phys.\ Rev.\ D {\bf 76} (2007) 064004. 

\bibitem{staro07}A.~A.~Starobinsky, JETP. Lett., {\bf 86} (2007) 157. 

\bibitem{odin07}S.~Nojiri and S.~D.~Odintsov, {\em Unifying inflation with LambdaCDM epoch in modified f(R) gravity consistent with Solar System tests}, arXiv:0707.1941 [hep-th]. 

\bibitem{AB}S.~A.~Appleby and R.~A.~Battye, arXiv:0705.3199[astro-ph]; L.~Pogosian and A.~Silvestri, arXiv:0709.0296[astro-ph]; S.~Tsujikawa, arXiv:0709.1391[astro-ph]; S.~Capozziello and S.~Tsujikawa, arXiv:0712.2268[gr-qc]. 

\bibitem{seba}G. Cognola, E. Elizalde, S. Nojiri, S.D. Odintsov, L. Sebastiani, S. Zerbini {\em Class of viable modified $f(R)$ gravities describing inflation and the onset of accelerated expansion} arkiv:0712.4017 [gr-qc]. 

\bibitem{cognola05}G. Cognola, E. Elizalde, S. Nojiri, S. D. Odintsov and S. Zerbini, JCAP {\bf 0502 }010 (2005). 

\bibitem{faraoni}V.~Faraoni, Ann.\ Phys.\ {\bf 317}, 366 (2005); V. Faraoni, Phys. Rev. {\bf D 72}, 061501 (2005) (R); V. Faraoni, Phys.\ Rev.\ D {\bf 75} (2007) 067302. 

\bibitem{cognola}G.~Cognola and S.~Zerbini, J.\ Phys.\ A {\bf 39} (2006) 6245. 

\bibitem{NO}S. Nojiri and S. D. Odintsov, Phys. Rev. {\bf D 68}123512 (2003). 

\bibitem{capo}S. Capozziello, S. Nojiri, S. D. Odintsov and A. Troisi, Phys. Letts. {\bf B 639} 135 (2006). 

\bibitem{NO1}S. Nojiri and S. D. Odintsov, Phys. Rev. {\bf D 74} 086009 (2006). 

\bibitem{NO2}S. Nojiri and S. D. Odintsov, J.\ Phys.\ Conf.\ Ser.\ {\bf 66} (2007) 012005. 

\bibitem{bazeia07}D. Bazeia, B.Carneiro da Cunha, R. Menezes, A.Yu. Petrov Phys.\ Lett.\ B {\bf 649} (2007) 445. 

\bibitem{rador}T.~Rador, Phys.\ Rev.\ D {\bf 75} (2007) 064033. T.~Rador, Phys.\ Lett.\ B {\bf 652} (2007) 228. 

\bibitem{soko}L.~M.~Sokolowski, Class.\ Quant.\ Grav.\ {\bf 24} (2007) 3713; L.~M.~Sokolowski, Class. Quantum Grav. 24 (2007) 3391-3411.

\bibitem{b}C.~G.~Bohmer, L.~Hollenstein and F.~S.~N.~Lobo, Phys.\ Rev.\ D {\bf 76} (2007) 084005. 

\bibitem{sasaki}S.~Nojiri, S.~D.~Odintsov and M.~Sasaki, Phys.\ Rev.\ D {\bf 71}, 123509 (2005). 

\bibitem{ART}I.~Antoniadis, J.~Rizos, K.~Tamvakis, Nucl.\ Phys.\ B {\bf 415}, 497 (1994); N.~E.~Mavromatos and J.~Rizos, Phys.\ Rev.\ D {\bf 62}, 124004 (2000); N.~E.~Mavromatos and J.~Rizos, Int.\ J.\ Mod.\ Phys.\ A {\bf 18}, 57 (2003). 

\bibitem{fGB}S.~Nojiri, S.~D. Odintsov, Phys.\ Lett.\ B {\bf 631} 1 (2005). 

\bibitem{Sami}M.~Sami, A.~Toporensky, P.~V.~Tretjakov and S.~Tsujikawa, Phys.\ Lett.\ B {\bf 619}, 193 (2005); S.~Tsujikawa and M.~Sami, JCAP {\bf 0701} (2007) 006. 

\bibitem{Mota}T.~Koivisto and D.~F.~Mota, Phys.\ Lett.\ B {\bf 644} (2007) 104; T.~Koivisto and D.~F.~Mota, Phys.\ Rev.\ D {\bf 75} (2007) 023518; Z.K.~Guo, N.~Ohta and S.~Tsujikawa, Phys.\ Rev.\ D {\bf 75} (2007) 023520; G. Calcagni, B. Carlos and A. De Felice, Nucl. Phys. B {\bf 752} 404 (2006). 

\bibitem{Calcagni}G.~Calcagni, S.~Tsujikawa and M.~Sami, Class.\ Quant.\ Grav.\ {\bf 22}, 3977 (2005); A.~Sanyal, Phys.\ Lett.\ B {\bf 645} (2007) 1. 

\bibitem{Neupane}I.~P.~Neupane, Class.\ Quant.\ Grav.\ {\bf 23} (2006) 7493; B.~M.~N.~Carter and I.~P.~Neupane, JCAP {\bf 0606} (2006) 004. 

\bibitem{GB}S.~Nojiri, S.~D.~Odintsov and O.~G.~Gorbunova, J.\ Phys.\ A {\bf 39} 6627(2006); I.~Brevik and J.~Quiroga, Int.\ J.\ Mod.\ Phys.\ D {\bf 16} (2007) 817. 

\bibitem{cognola06}G. Cognola, E. Elizalde, S. Nojiri, S.D. Odintsov and S. Zerbini, Phys. Rev. {\bf D 75}086002 (2007). 

\bibitem{cognola066}G. Cognola, E. Elizalde, S. Nojiri, S. D. Odintsov and S. Zerbini, Phys. Rev. {\bf D 73} 084007 (2006). 

\bibitem{sami}S.~Nojiri, S.~D.~Odintsov and M.~Sami, Phys.~Rev.~D {\bf 74} (2006) 046004. 

\bibitem{E}E.~Elizalde, S.~Jhingan, S.~Nojiri, S.~D.~Odintsov, M.~Sami and I.~Thongkool, {\em Dark energy generated from a (super)string effective action with higher order curvature corrections and a dynamical dilaton}, arXiv:0705.1211 [hep-th]; S.~Nojiri, S.~D.~Odintsov and P.~V.~Tretyakov, Phys.\ Lett.\ B {\bf 651} (2007) 224. 

\bibitem{vilenkin85}A. Vilenkin, {\sl Phys. Rev.} {\bf D 32} (1985) 2511. 

\bibitem{capozziello02}S. Capozziello, Int. J. Mod. Phys. {\bf D11}4483 (2002). 

\bibitem{cognolaV}G. Cognola, S. Zerbini, Vestnik TSPU, {\bf 7}, 69 (2004). 

\bibitem{staro}A. Starobinsky, {\sl Phys.Lett.} {\bf B91} (1980) 99. 

\bibitem{staro1}A. Starobinsky and H. J. Schmidt, {\sl Class. Quantum Grav.} {\bf 4} 695 (1987); V. Muller, H. J. Schmidt and A. Starobinsky, {\sl Phys.Lett.} {\bf B 202} 198 (1988); H. J. Schmidt, {\sl Class. Quantum Grav.} {\bf 5} 233 (1988) 

\bibitem{aco}I. Navarro and K. Van Acoleyen, JCAP {\bf 0603} 008 (2006); T. Chiba, JCAP {\bf 0503} 008 (2005); A. Nunez and S. Solganik, Phys. Letters {\bf B 608} 189 (2005). 

\bibitem{herv}J. D. Barrow and S. Hervik, Phys. Rev. {\bf D 74}124017 (2006); B.~Li, J.~D.~Barrow and D.~F.~Mota, Phys.\ Rev.\ D {\bf 76} (2007) 044027; B.~Li, J.~D.~Barrow and D.~F.~Mota, Phys. Rev. D 76, 104047 (2007). 

\bibitem{topo}A.V. Toporensky, P.V. Tretyakov, Int.\ J.\ Mod.\ Phys.\ D {\bf 16} (2007) 10751086. 

\bibitem{waldram}A. Hindawi, B. A. Ovrut, and D. Waldram, Phys. Rev. {\bf D 53} 5597, (1996). 

\bibitem{sagno}M. H.Goroff and A. Sagnotti, {\sl Nucl. Phys.} {\bf B 266}709 (1986). 

\bibitem{van}A.E.M. van den Ven, {\sl Nucl. Phys.} {\bf B 378}309 (1992).   


\end{thebibliography}
\end{document}